# Design of a dynamic and self-adapting system, supported with artificial intelligence, machine learning and real-time intelligence for predictive cyber risk analytics in extreme environments – cyber risk in the colonisation of Mars

Petar Radanliev[1] · David De Roure[1] · Kevin Page[1] · Max Van Kleek[2] · Omar Santos[3] · La'Treall Maddox[3] · Pete Burnap[4] · Eirini Anthi[4] · Carsten Maple[5]

## Abstract
Multiple governmental agencies and private organisations have made commitments for the colonisation of Mars. Such colonisation requires complex systems and infrastructure that could be very costly to repair or replace in cases of cyber-attacks. This paper surveys deep learning algorithms, IoT cyber security and risk models, and established mathematical formulas to identify the best approach for developing a dynamic and self-adapting system for predictive cyber risk analytics supported with Artificial Intelligence and Machine Learning and real-time intelligence in edge computing. The paper presents a new mathematical approach for integrating concepts for cognition engine design, edge computing and Artificial Intelligence and Machine Learning to automate anomaly detection. This engine instigates a step change by applying Artificial Intelligence and Machine Learning embedded at the edge of IoT networks, to deliver safe and functional real-time intelligence for predictive cyber risk analytics. This will enhance capacities for risk analytics and assists in the creation of a comprehensive and systematic understanding of the opportunities and threats that arise when edge computing nodes are deployed, and when Artificial Intelligence and Machine Learning technologies are migrated to the periphery of the internet and into local IoT networks.

## Introduction

The advancements of artificial intelligence in industrial automation, triggers questions on safety and security, and whether AI has enhanced security or increased the cyber risks in extreme environments (Khan 2020), This is specifically of concern in some extreme environments where cyber-attacks can cause significant and irreversible damage, such as the space

✉ Petar Radanliev
petar.radanliev@oerc.ox.ac.uk

David De Roure
david.de.roure@oerc.ox.ac.uk

Max Van Kleek
m.d.c@oxford.ac.uk

Omar Santos
oma.rsantos@cisco.com

La'Treall Maddox
latrea.llmadox@cisco.com

Pete Burnap
p.b.urnap@cs.cardiff.ac.uk

Eirini Anthi
e.ant@cs.cardiff.ac.uk

Carsten Maple
k.m.c@warwic.ac.uk

[1] Department of Engineering Sciences, University of Oxford, Oxford e-Research Centre, 7 Keble Road, Oxford OX1 3QG, UK

[2] Department of Computer Science, University of Oxford, Oxford, UK

[3] Cisco Research Centre, Research Triangle Park, Raleigh-Durham, NC, USA

[4] School of Computer Science and Informatics, Cardiff University, Cardiff, UK

[5] MG Cyber Security Centre, University of Warwick, Coventry, UK

Springer

exploration industry. Commitments on the colonisation of Mars have emerged from multiple governmental agencies (e.g. NASA, ESA, Roscosmos, ISRO and the CNSA) and private organisations (e.g. SpaceX, Lockheed Martin, and Boeing). The colonisation of Mars presents many difficulties and hazards, such as radiation exposure, toxic soil, low gravity, lack of water, cold temperatures, and social isolation. One specific risk has been ignored until now, the risk of cyber-attacks. Since the colonisation of space requires the reinstalment of critical infrastructure, and such infrastructure needs to be installed by smart machines, then we need to anticipate that the risk assessment of such smart machines operating autonomously, should be undertaken by other smart machines. We also need to anticipate the cost (including the cost of down time) required for repairing or replacing such machine – in outer space. This cost (and cost of down time) should be understood prior to deciding on the most appropriate space systems. Although such systems can be tested prior to deployment, the risk assessment after deployment would be completely reliant on Artificial Intelligence (AI), because there will be no human participation in the installation phases.

In this paper, we conduct a survey review of different Artificial Intelligence and Machine Learning (AI/ML) methods, that could be used for the risk assessment after deployment. Recent studies on Artificial Intelligence and Machine Learning (AI/ML) perspectives on mobile edge computing lack detail (Radanliev et al. 2020d), but provide guidance on how data can be processed in real-time, reducing edge-cloud delay and inform on the topic of cognitive cyber security at the edge. Since the risk assessment of deployed space systems is extremely difficult, due to the extreme conditions, this paper is focused on the topic of predicting cyber risk loss magnitude through dynamic analytics of cyber-attack threat event frequencies. Forecasting the threat events, could enable preventing such events from occurring in the first place. Secondly, we investigate what data is required for applying AI algorithms in dynamic risk analytics. Additional challenges addressed in this study relate mainly to socio-technical issues, such as technology, regulation, supply chains and control systems. For example, investigating the perceptions of risk and trustworthiness that emerge as a result of machine agency, which interact with regulation, standards and policy on the one hand and design and engineering on the other, spanning the physical and behavioural sciences. But the specific focus of this paper is on integrating AI/ML in the data collection and analytics of risk through fog computing (i.e. use of edge devices) for forward-facing predictive outputs. We investigate a scenario where an organisation planning for colonisation of outer space, has implemented all the security recommendations (e.g. NIST), but the risk remains from uncertain and unpredictive attack vectors in outer space, at the edge of the network.

For narrowing the topic to assessment of these new types of cyber security, the research adopts a red teaming methodology for detecting and reducing threats and simplify compliance with internal, industry and government regulations. A red teaming approach is firstly applied by challenging plans, policies, systems and assumptions and adopting an adversarial approach to IoT cyber risks. With this approach, IoT cyber risks can be divided in three levels, edge, fog and cloud. The fog computing is placed in the distribution network layer and provides sufficient computational resources, low latency and compute-intensive applications. The cloud computing level represent a shared pool of rapidly provisioned computing resources, for high computation and data storage. Hence, IoT cybersecurity deployment in the fog and cloud computing levels is not a big concern. The small computation capability at the edge devices makes IoT cyber risk more likely to occur at the edge computing level. Hence, this article is primarily focused on the edge computing level.

Since the focus of this review was the risk at the edge of the network, we applied this as the basis of selecting the AI methodologies considered in Table 1. However, there are many techniques that are not discussed in this review. A holistic review of all AI methodologies was considered beyond the scope of this review, and since many AI methodologies are not directly relevant to this topic, we selected only the most appropriate AI methodologies. To select the most appropriate AI methodologies, we applied a red teaming approach to identify IoT systems that are mostly affected by a few types of network risk event. Those include: Eavesdropping Attacks, Denial of Service (DoS) and Distributed DoS (DDoS), Spoofing Attacks, and Man-in-the-Middle attacks (MITM). To describe briefly the relationship between these types of attacks, Eavesdropping Attacks is used for listening IoT communications without the transmission appearing abnormal, hence making it difficult to detect. After Eavesdropping Attacks has gained authorisation access, Spoofing Attacks are used to send spoofed traffic with a legitimate access to IoT network. The MITM is just an advanced Spoofing Attack where adversary is positioned between two IoT devices and independently intercepts data and communicates between endpoints, collecting sensitive information, dropping packets, and causing different security vulnerabilities. The DoS and DDoS floods the IoT devices network with traffic, this overloads the communication and exhausts the network, leading to IoT devices being unable to communicate. As simple as it is, this is the most common and most dangerous IoT attack. The small computation capability at the edge devices, make DDoS attacks really difficult to resolve. While new cyber security is constantly been developed (e.g. ISO 3000), the level of cyber-attack sophistication is also increasing (NetScouts 2018) (e.g. the Mirai variants 'VPN filter' is delivered in multiple stages with modularised payload; 'TORii' uses its own encryption and evasion tactic). Considering these continuous changes, to assess the effectiveness of cybersecurity, we need cyber analytic approaches that can handle real time intelligence in the form of probabilistic data collected at the edge. But the effectiveness of cybersecurity should not only be measured by the protection of cyberspace, but also with the protection of assets that can be reached via cyberspace (Davis et al. 2019).

Table 1  AI/ML algorithm application for descriptive, predictive, and prescriptive risk analytics in edge computing

| AI/ML technique | Application | References |
| --- | --- | --- |
| Deep learning – ANN | Network architecture | Berman et al. (2019), Diro and Chilamkurti (2018), Roopak et al. (2019), Vinayakumar et al. (2019) |
| Anomaly detection, unsupervised learning, classification | Network planning/load balancing | Gebremariam et al. (2019), Ullah et al. (2019) |
| Regression | | Hu et al. (2017) |
| Classification – Bayesian networks | Fault and failure detection/management | Bashir et al. (2019), Sultana et al. (2019) |
| Classification/Clustering – autoencoders | | Al-Turjman (2020) |
| Algorithms: supervised, unsupervised and reinforcement learning | Network management and operations | Cui et al. (2019), Nguyen et al. (2019) |
| ANN (RNN) and random forest | | Park et al. (2018) |
| Regression – ANN | | Anagnostopoulos and Hadjiefthymiades (2019) |
| Classification – Naive Bayes (NB) | | (Syafrudin et al. (2018), Yin et al. (2019) |
| Classification algorithms: K nearest neighbours, SVM | | Guo et al. (2018), Sangaiah et al. (2019), Zhang et al. (2019) |
| GDNN | | J. Wang et al. (2019a), X. Wang et al. (2019b) |
| ANN | Network security and breach detection | Sun et al. (2019) |
| Dynamic game – Nash Folk theorem | | Abegunde et al. (2016) |
| Game theory and NB classifier | | Bui et al. (2017), Moura and Hutchison (2019) |
| Deep learning algorithms | | Blanco-Filgueira et al. (2019), Li et al. (2018) |
| Algorithms: supervised, unsupervised and reinforcement learning | | Cao et al. (2019) |

In brief, we investigate the role of AI/ML in cyber risk analytics with use of confidence intervals and time bound ranges at the edge. The objective of such an approach would be to protect data integrity, while securing predictive analytic outputs and integrating such solutions in these new types of edge computing cyber security. In edge computing, the IoT-augmented physical reality is open to adversarial behaviours that are yet uncharted and poorly understood, especially the socio-technical dimensions. This paper evaluates the impact of compromise in terms of its safety implications and resulting consequences on end system provision.

# Research methodology

The research methodology applied consist of a survey review investigating different deep learning and machine learning algorithms and their application in AI for securing the edge. The survey review is used for investigating the intersections between cyber risk and technology, regulatory interventions, and economics.

The research methodology in this was survey review paper, was developed to address the (un)availability of data. Although there is a valid concern about the (un)availability of data, in the present digital age, the IT and IoT devices create a large volume of data. Hence, the real challenges that remain, are in developing suitable data strategies to utilise this new data. Simultaneously, the cyber security architecture for complex coupled systems, demands data strategy optimisation and decision making on collecting and assessment of probabilistic data. With consideration of the above, the research integrates impact assessment models, with AI and risk analytics models, for developing a dynamic and self-adopting data analytics methodology to assess, predict, and analyse cyber-risks.

For risk assessment of systems for colonising outer space, we need a quantitative risk impact estimation is needed – for estimating cyber security and cyber risk at the edge (Radanliev et al. 2020e). Our argument is that without a dynamic real-time probabilistic risk data and cyber risk analytics enhanced with AI/ML, these estimations can be outdated and imprecise. We are concerned not just with securing a system, but to acknowledge that failure and compromise will occur and address how the system responds in these circumstances. This is an important methodological principle which distinguishes out work within the cybersecurity domain. Recent literature confirms diverse cyber risks from IoT systems (Maple 2017), including risks in IoT ecosystems (Tanczer et al. 2018) and IoT environments (Breza et al. 2018), such as risk from smart homes (Eirini Anthi et al. 2019; Ghirardello et al. 2018), the Industrial IoT (Boyes et al. 2018), and challenges in security metrics (Agyepong et al. 2019). Cybersecurity solutions for specific IoT risks are also emerging at a fast rate, such as new models on opportunities and motivations for reducing cyber risk (Safa et al. 2018), adaptive intrusion detection (E. Anthi et al. 2018), security economic by design (Craggs and Rashid 2017), highlighting the privacy requirements (Anthonysamy et al. 2017) and strategies for achieving privacy (Van Kleek et al. 2018). Therefore, our methodology is based on mathematical principles and quantitative data. In recent publications on this topic (Radanliev et al. 2020b), we discovered that the lack of probabilistic data leads to qualitative cyber risk assessment approaches, where the outcome represents a speculative assumption. Emerging quantitative models are effectively designed with ranges and confidence intervals based on expert opinions and not probabilistic data (Buith 2016).

## Survey of AI/ML algorithms

The AI/ML are essential for advancing beyond the limitations of Value-at-Risk (VaR) models (FAIR 2017), where Bayesian and frequentist methods are applied with and beyond VaR models (Malhotra 2018). This requires federated learning and blockchain based decentralised AI architecture where AI processing shifts from the cloud to the edge and the AI workflow is moved and data restricted to the device (Porambage et al. 2019). Current gaps in cyber risk analytics are in the areas of descriptive, predictive, and prescriptive analytics (Barker et al. 2017). Hence, a survey of AI/ML applications is presented in Table 1, to address the main questions emerging from this study on edge computing and descriptive, predictive, and prescriptive risk analytics.

Table 1 confirms that by integrating AI/ML in the risk analytics, we can devise a new approach for cognitive data analytics, creating a stronger resilience of systems through cognition in their physical, digital and social dimensions. This approach resolves around understanding how and when compromises happen, to enable systems to adapt and continue to operate safely and securely when they have been compromised. Cognition through AI/ML and how cognitive real time intelligence would enable systems to recover and become more robust is evaluated in more detail below. The survey in Table 1 is informed by but avoids overlapping with a series of working papers and project reports on IoT cyber risk, IoT risk assessment and IoT at the edge found in pre-prints online. This research is specifically focused on AI/ML in IoT risk analytics, and it benefits from this established research knowledge. But with a focus on the topic of securing the edge through AI/ML real time analytics to build stronger transformative and impactful understanding on the topic.

Majority of the current Intrusion Detection Systems (IDS) are based on ML algorithms and the CNN (Convolutional Neural Network) + LSTM (Long Short Term Memory) appear to perform better than other Deep Learning (subsets of ML) models (Roopak et al. 2019). Such arguments are difficult to generalise when tested with a single dataset. Deep Neural Network (DNN) has been applied with distributed deep learning to collect network-based and host-based intrusion detection systems (NBID and HBID) (Vinayakumar et al. 2019). This is a very comprehensive study, where a multilayer perceptron (MLP) model is adopted. However, in a related research, the MLP (type of artificial neural network - ANN) was found to be the least accurate deep learning model (Roopak et al. 2019).

Network Based Intrusion Detection Systems (NIDS) that use statistical measures or computer thresholds have been related to security research since the early days of computer architecture (Vinayakumar et al. 2019). But are ineffective for current cyber risk analytics of connected and highly complex ICT systems, because they present high rates of false negatives (failure to detect) and false positives (false alerts). Distributed attack detection at fog level was proven to be more scalable than centralised cloud for IoT (Diro and Chilamkurti 2018). If the attack vectors are known, then up to 99.999% accuracy can be reached by type of attack with bidirectional long short-term memory (LSTM) units introduced to recurrent neural network (RNN) (Berman et al. 2019). Similarly, a Siamese Network Classification Framework (SNCF) can alleviate imbalance in risk prediction and present more reliable results when compared with other algorithms (Sun et al. 2019). With SNCF two different types of risk data sets can be used, (1) public data set (less features and more samples), (2) real data set (more features and less samples). The first set could verify solving the imbalance problem, and the second could eliminate reliance on the characteristics of feature engineering. Such experimental SNCF results have shown good cyber risk prediction performance (Sun et al. 2019) and Software Defined Networking Technology (SDN) has been effective in detecting and monitoring network security when integrated with Machine Learning (ML) and deep learning (DL) to create SDN-based NIDS (Sultana et al. 2019). The main risk concern with SDN and Network Functions Virtualisation (NFV) is the centralised nature which creates a single point of failure (Gebremariam et al. 2019). To resolve this, three layered nodes (Edge-IDS, Fog-IDS, and Cloud-IDS) has been proposed for NIDS system in SDN-based cloud IoT networks (Nguyen et al. 2019). Cloud environments enable IoT device virtualisation resulting with virtual IoT objects that can be accessed and controlled remotely though a dynamic virtual network (Ullah et al. 2019).

A power load forecasting (Hu et al. 2017), can be based on the generalised regression neural network with decreasing step fruit fly optimisation algorithm. Similarly, logistic regression and multicriteria decision making in IoT fog computing can be used for resource allocation (Bashir et al. 2019). The main concern we have about the development of such algorithms is that deployment of 5G can separate real-time intelligence and security between IoT, IoE or even IoNT (Al-Turjman 2020). Hence, intelligence and cognition techniques would differ in application areas and architecture. One of the possible issues is that ML platforms (such as TensorFlow, Gaia, Petuum, Apache Spark, and GraphLab), are designed for offline data analytics and training data are collected, partitioned, and learned offline to construct machines for data analytics (Cui et al. 2019). While some of the recently proposed detection systems for edge computing are operating in real time, e.g. LiReD (Park et al. 2018). Edge nodes can host and process the data to limit latency, and recently enhanced models can handle the earlier problems with missing values (Anagnostopoulos and Hadjiefthymiades 2019), while improving the detection accuracy (Yin et al. 2019) and decision making with early warning systems (Syafrudin et al. 2018). The classification accuracy can also be improved with edge filtering (Guo et al. 2018), position confidentiality (Sangaiah et al. 2019) and dynamic data classification (Zhang et al. 2019), to avoid system overload when tasks increase suddenly, by diverting and allocating complex tasks to devices with stronger computing power.

Multi-Access Edge Computing based on reinforcement learning, enhances the performance of such 'offloading' in polynomial time complexity - worst-case running time (J. Wang et al. 2019a). While integration of Deep Reinforcement Learning and Federated Learning with mobile edge systems, optimises mobile edge computing, caching, and communication, and makes edge systems more intelligent (X. Wang et al. 2019b). Optimising and balancing resource constrains in edge computing has been investigated with 'dynamic game'(Abegunde et al. 2016) and 'game theory' (Moura and Hutchison 2019) strategies. Such optimisation is primarily theoretical, but highly relevant for red teaming of edge computing risks. Two models 'Cournot' and 'Stackelberg' are proposed for making real-time optimisation of traffic flow (Bui et al. 2017). These models need to be tested with real-time data to be verified, but the theoretical contribution is quite significant, e.g. applying the 'Prisoners Dilemma' on optimising decisions.

Deep learning models recorded highest accuracy as 97.16% detection of DDoS attacks (Roopak et al. 2019), and the multi-layered structure, makes them very adoptable to edge computing. Hence, deep learning has been applied for optimising performance while protecting user privacy in uploading data, (Li et al. 2018). But the computing and memory requirements, along with the high power consumption, make them difficult to use in edge computing (Blanco-Filgueira et al. 2019). Further research is needed to identify how deep learning can be applied in practice, with real-time data. Possibly reinforcement learning, supervised/unsupervised learning, and deep reinforcement learning (Cao et al. 2019), would provide some insights into how this can be achieved.

## Elements of artificial intelligence and machine learning in cognition engine design

Cyber risk analytics at present is reactive and assessments are based on risk/loss events that already occurred (Radanliev et al. 2020a). AI/ML in forward-looking predictive analytics enable threat intelligence prediction and faster attack detection. The main advantage of AI in risk analytics is the fast processing and analysis of big data where parsing, filtering and visualisation is done in near real time (Radanliev et al. 2020c). Machine learning uses mathematical and statistical methods and algorithms that learn, build and improve models from data. This enables design of a cognition engine in the form of automated predictive cyber intelligent software agents that identify, assess and record cyber-attacks. After this, natural language processing (NLP) can be applied to perform behaviour analytics and create baseline profiles of normal behaviour and then monitor for abnormalities while continuously learning from the profile's behaviour patterns. Facilitating a consistent and repeatable detection of threat indicators and predictions about new persistent risks that are undetected. AI/ML learn from multiple patterns (e.g. threat intelligence feed, device event logs, vulnerability information, contextual data) to determine predictive risk insights. Predictive risk analytics for advance notice of risk exposure and potential loss can be performed through monitoring the risk lifecycle activities, e.g. the reactive activities that capture losses and near miss events. From reactive activities we can quantify the impact of losses and develop baseline indicators to compare mathematical results.

### Mathematical formulae

To develop predictive risk analytic methodology for estimating the loss of cyber risk, we apply adapted version of the aggregate loss method to compound a Poisson discrete probability distribution. For the adopted version, we use the theoretical cumulative distribution function of aggregate loss, as shown in (Charpentier 2014):

$$F_{L_c}(l) = \sum_{n=0}^{+\infty} P(N = n) P\left(\sum_{i=1}^{n} Z_i < l\right) \quad (1)$$

In the adapted version, we generate the frequency distribution from the cumulative function in Eq. 1, with non-linear summation and simulated random variables to approximate the theoretical function.

The $L\mathscr{C}$ = aggregate loss distribution consisting of the compound sum of $N$ = frequency (intensity) and $Z\mathscr{I}$ = severity (loss) distribution and is described as: $L\mathscr{C} = \sum_{i=1}^{N} Z\mathscr{I}$, where $L\mathscr{C} = 0$, if $N = 0$.

Considering the (un)availability of probabilistic data, the $N$, $Z\mathscr{I}$, and the consequent $Zj$ where $(\mathscr{I}j)$ are considered independent. This cumulative function defines a frequency distribution for aggregate loss as nonlinear summation. The function can be improved by considering the frequency distribution as Poisson variable, where for a given time interval $[0,t_c]$, the inter-arrival time = $S_i$ of two 'risks' within the interval follows an exponential distribution with parameter $\lambda_c$. This function can be described as:

$$P(N(t) = n) = \frac{\exp(-\lambda_c t_c)(\lambda_c t_c)}{n!} \quad (2)$$

$$Y_i = S_i - S_{i-1} \sim Exp(\lambda_c), F(Y < y) = 1 - \exp\{-\lambda_c y\} \quad (3)$$

The known issues with (un)availability of sufficient probabilistic data (Radanliev et al. 2018) can be mitigated by enhancing the precision of the sample size in the inter-arrival time, where the insufficient (few years) data can be considered as lognormal (Galton) distribution where $t_c = 365$ (representing 365 days). In a more specific dataset scenarios, the distribution will vary depending on the probabilistic data. We postulate that the $t_c = 365$ has a fixed loss per day = $b$, where $M_x$ = total loss days for an IoT device $IoT_x$ and the device is operational at time $t$ and the total loss per $t_c = b \times M_x$. Considering that IoT device can stop functioning (or be killed by grey-hat attack) at any point in $t_c$, then $M_x$ represents a continuous random variable of the future (potential) loss from an IoT device infected for time $x$ and $1_{(\cdot)}$, with a given discount rate = $r$ calculated as $v = (1+r)^{-1}$, then considering the probability of IoT device stops (or be killed) and the discount factor, the present potential loss $P_x$ can be determined as:

$$P_x = v \times 1_{(T_x > 1)} \times (b \times M_x) \quad (4)$$

This formula calculates the risk of loss depending on the IoT device surviving the entire 365 days, or stops (or be killed) during the 365 days. The second postulate is that the risk of loss is eliminated when the device is killed. The actuarial equivalent of this can be explained as the present values of the expected losses described as loading = $\delta$ and expected revenue (that was lost) = $\pi_\chi$ are equal to:

$$\pi_\chi = (1+\delta) \times E(P_x) \quad (5)$$

In time, when more extensive data from IoT devices becomes available, more precise $\delta$ can result with lowering the expected loss = $\propto$ and a more precise expected present value of the loss = $E\left(P_\chi^1\right)$ can be estimated as:

$$E\left(P_\chi^1\right) = \propto \times E(P_x) \quad (6)$$

and expected revenue (that was lost) as:

$$P_\chi^1 = (1+\delta) \times E\left(P_\chi^1\right) \quad (7)$$

The continuous random variable of the future (potential) loss $M_x$ can be divided on the number of attacks (frequency) = $N_x$ and the and severity (loss) = $Z\mathscr{I}$ per breach in a given $t_c$ can be denoted as $R_{x,i}$ and $N_x$ would reflect a Poisson distribution with time-varying intensity $\theta_\chi$, and $R_{\chi,0}$ and $R_{\chi,1} = \Upsilon_{\chi,1} + 1$ where $\Upsilon_{\chi,1}$ follows time-varying intensity = $\lambda_\chi$. The $M_x$ in a given $t_c$ for an IoT device $IoT_x$ can be estimated as:

$$M_x = \sum_{i=0}^{N_\chi} R_{x,i} \quad (8)$$

With a compound Poisson process, the probable present potential loss $P_x$ with a given $M_x$ in a given $t_c$ for an IoT device $IoT_x$, can be calculated with:

$$P(M_x = n) = \begin{cases} \exp(-\theta_\chi), n = 0 \\ \sum_{j=1}^{n} \frac{j(\lambda_\chi)^{n-j}(\theta_\chi)^j \exp-(j\lambda_\chi + \theta_\chi)}{j!(n-j)!}, \\ n \geq 1 \end{cases} \quad (9)$$

The above equation is designed for IoT risk assessment, but it can easily be adopted for different types of cyber risks. For example, we could calculate IoT cyber risk from AI as $\kappa$ for a given IoT device $IoT_x$ with $\kappa \times \pi_\chi^1$, where in the first instance, the total loss $L_\kappa$ would include $M_x$ and $L\mathscr{C}$. This can be expressed as:

$$L_\kappa = \sum_{j=1}^{k} M_x + L\mathscr{C} \quad (10)$$

and evaluated with risk proxies from shortfall probability, Value at Risk and Conditional Tail Expectation. The shortfall probability can be calculated as:

$$Prob(Shortfall) = Prob\left(\kappa \times \pi_\chi^1 \leq L_\kappa\right) \quad (11)$$

where expected shortfall is:

$$E(Shortfall) = E\left(max\left(\kappa \times \pi_\chi^1 - L_\kappa, \ 0\right)\right) \quad (12)$$

With $r$ (described earlier), and the threshold $=\rho$, Value at Risk can be calculated as:

$$VAR(\rho) = \inf\left\{L_\kappa^* | F_{L_K}\left(L_\kappa^*\right) < 1-\rho\%\right\} \quad (13)$$

and the:

$$CTE(\rho) = E(L_\kappa | L_\kappa \geq VaR(\rho)) \quad (14)$$

With the VaR and CTE risk proxies, we can calculate the risk margin ratio $= \delta(L)$ with the Solvency 2 Directive percentile method:

$$\delta(L) = \frac{\rho(L) - E(L)}{E(L)} \quad (15)$$

where $\rho(L)$ represents VaR and CTE risk measures, and $E(L)$ the best estimate. If this is considered with an assumption that losses would be larger than ransoms: $\propto$ of losses $\leq \propto$ of ransoms. Then the power-law distribution can be calculated with the equation:

$$P(x) = \propto X^{-\propto} \quad (16)$$

where the variance analysis of $\propto$ parameter is $1 < \propto < 2$, with infinite mean and average even when $2 < \propto < 3$.

## Cognitive design

Connecting the lost exposure of cyber risk from human-computer interaction (frequency), in different information knowledge management systems (magnitude), with artificial intelligence, can provide predictive feedback sensors for primary and secondary loss (vulnerabilities). These feedback sensors represent dynamic real time data mechanisms that assist and enable better understanding of the vulnerabilities - prior to cyber-attacks. The reliability of cyber risk analytics could increase significantly if decisionmakers have a dynamic and self-adopting AI enhanced feedback sensors to assess, predict, analyse and address the economic risks of cyber-attacks.

The survey (in Table 1) identified all relevant AI algorithms, and the mathematical formulae (results in Table 2) articulates some of the possible solutions for the role of these algorithms in designing dynamic automated predictive feedback cognitive system, supported with real-time intelligence.

Cyber risk analytic approaches with dynamic real-time and AI/ML self-adapting enhanced technologies that enable predictive risk analytics are identified in Table 1. While the design of a predictive cyber risk analytics is based on confidence intervals and time bound ranges in Table 2. In doing this work we are acutely aware that adding automation and further coupling to a distributed system also brings new opportunities for cascading effects and exposing new attack surfaces. These concerns are fundamental in the areas with increased automation of processes which have classically required human interaction.

## Dynamic and self-adapting predictive data analytics with the mathematical formulae

A range of data sources was used to apply data analytics with the new mathematical formulae. The Comprehensive Threat Intelligence was used to collect data from vulnerability reports and zero-day reports (Cisco 2020). The Chronology of Data Breaches (Land et al. 2020) was used to gather larger sample size from thousands of records collected over the last 10 years (2010–2020). The SonicWall cyber threat report was used to collect probabilistic data on trends of IoT attacks (SonicWall 2019). The aggregate cyber risk from a large sample population is not the ideal measure for calculating the cyber risk of a small and/or medium sized enterprise. Hence, we divided the large sample into subsamples that follow a Poisson distribution with smaller total risk $\lambda_c$, where:

$$\lambda_c = \lambda_{c1} + \ldots + \lambda_{cm} \quad (17)$$

and $\lambda_{ci}, i = 1, \ldots m$ represented as the individual risk of a subsample. Finally, the total cyber risk of the adjusted proportion parameter $p$ is equal to $\lambda_c = p\lambda_c$. We estimate risk exposure of total IoT cyber risk $p_1$ and the IoT cyber risk from non-recorded devices as $p_2$ where:

**Table 2** Dynamic and self-adapting predictive cyber risk analytics based on different levels of cyber risk intelligence

| Risk calculation metrics | Cyber risk | IoT cyber risk | | | |
|---|---|---|---|---|---|
| | | Guarded (Green) | Elevated (Yellow) | High (Amber) | Severe (Red) |
| $E\left(P_\chi^1\right)$ | 9,225,798 | 8,302,872 | 8,323,645 | 8,495,883 | 8,826,248.5 |
| $Prob\left(\kappa \times \pi_\chi^1 \leq L_\kappa\right)$ | 0.362% | 0.390% | 0.717% | 2.952% | 4.359% |
| $E\left(max\left(\kappa \times \pi_\chi^1 - L_\kappa, 0\right)\right)$ | 776 | 783 | 31,660 | 281,340 | 807,491 |
| VAR(.90) | 9,659,815 | 8,696,453.5 | 8,706,061 | 8,797,711.5 | 8,947,727 |
| VAR(.95) | 9,785,002.5 | 8,807,375.5 | 8,823,096.5 | 9,020,017.5 | 9,510,794.5 |
| VAR(.99) | 10,020,820.5 | 9,031,839.5 | 9,092,977.5 | 11,314,474.5 | 14,126,426.5 |
| CTE(.90) | 9,823,671.5 | 8,844,958.5 | 9,021,798 | 10,482,011.5 | 13,461,685 |
| CTE(.95) | 9,930,712 | 8,942,115.5 | 9,283,971 | 12,076,385 | 17,770,438 |
| CTE(.99) | 10,139,834 | 9,137,848.5 | 10,729,379.5 | 22,130,721.5 | 45,991,022 |

$E\left(P_\chi^1\right)$ = Expected present value of the loss;

$Prob\left(\kappa \times \pi_\chi^1 \leq L_\kappa\right)$ = Prob(Shortfall) or Shortfall probability;

$E\left(max\left(\kappa \times \pi_\chi^1 - L_\kappa, 0\right)\right)$ = E(Shortfall) or Shortfall of expected present value of the loss;

VAR($\rho$) = Value at Risk;

CTE($\rho$) = Conditional Tail Expectation

Note: Assuming N = 1000, b = 1000, r = 0.03, ∝ = 0.9, δ = 0.1 in the IoT cyber risk calculation and number of repetitions = 100,000

$$p = p_1 \times \left(m \times p_2^{-1}\right); \quad (18)$$

To demonstrate how these models can be applied for numerical results, in this section we present a demonstration project of different numerical results. For the numerical estimates in Table 2, we generate 100,000 Monte Carlo simulation runs, for 10,000 IoT devices being hacked. We use these Monte Carlo simulation runs to compare the effectiveness of different risk measures to quantify cyber risk, using the estimate of aggregate distribution of total loses (primary and secondary). We assume that by understanding the cyber risk, we can lower the frequency of cyber breaches (∝ = 0.9). In the next paragraph, we detail our data sources, and how we obtained the required data to present the demonstration of numerical results in Table 2. In the demonstration project, we adopted the numerical results to compare the estimated risk measures under different risk levels: Guarded (Green); Elevated (Yellow); High (Amber); Severe (Red).

$m$ = independent sample size. If we assume that $m$= 10,000 and $p$ = 0.00002. The $p$ = 0.00002 derived from the findings that IoT devices are attacked within 5 min of being connected to the internet (NetScouts 2018), while over 50% of the cyber risk professionals do not keep inventory of IoT devices installed (SFG 2017), hence there are potentially over 50% more IoT devices exposed to attacks. This is calculated as 365 (days in a year) × 24 (hours per day) × 60 (minutes per hour) – 50% (the cyber-attacks on not recorded IoT devices)[1] where ten times $p$ = 'high (amber)' risk, twenty times $p$ = 'severe (red)' risk. This reflects on findings that IoT will increase at a rate of 152,200 devices per minute by 2025 (Rosen 2015) × 525,600 (minutes in 365 days) = 80billion new IoT devices connected annually. This will increase the overall IoT cyber risk level. The twenty times assumption is based on the SonicWall report (SonicWall 2019) stating that IoT malware attacks increased by 215.7% from 10.3 m in 2017 to 32.7 m in 2018 and the trend continued in 2019.[2] The twenty times assumption represents .99 in Table 1. The corresponding .95 and .90 derive from the .99 calculation. We can also realistically assume that 'guarded (green)' level of cyber maturity would lower the ∝ = .90, then we can calculate the shortfall probability, expected shortfall, VaR and CTE for different cyber risk levels and tail risk under different assumptions in Table 1.

While existing cyber risk assessment models are based on individual risk calculation metrics, the approach presented in the mathematical formulae and demonstrated in Table 1, is based on multiple numerical risk metrics. The quantitative approach of the mathematical formulae, when integrated with Excel Macros, presents risk categorisations (Table 1) that are supported with real time intelligence. This presents a dynamic and self-adopting predictive cyber risk analytics approach, that is compliant with the existing NIST 'traffic lights' risk categorisations. The quantitative approach also correlates the NIST standards with the FAIR Institute efforts for quantitative cyber risk analytics. The mathematical formulae is similar to the FAIR-U approach (FAIR 2020), but instead of relying on a specific risk metric, its reliant on multiple numerical risk metrics. For comparison, the mathematical formulae uses different tail risk measurement and compares the impact of cyber risk under different risk categories. The 'high (amber)' and 'severe (red)' risk categories derive numerical representation of how rare and extreme events (black swan events) can increase the cyber risk impact. The impact of risk in VaR (.90) and CTE (.90) is not significant, but the risk margin ration increases significantly when compared to VaR (.99) and CTE (.99). This provides a quantitative perspective of impact from 'black swan' events, and enables more informative decision making on implementation of low cost and low security vs higher cost and higher security IoT systems, while putting 'black swan' events in IoT risk perspective. Worth noting that although IoT devices today are attacked within 5 min of being connected to the internet (NetScouts 2018), computers connected to internet even back in 2007 were attacked on average every 39 s (Cukier 2007). Given that computers even in 2007 had much more computing power than most IoT devices today, we can anticipate a continuous increase in attack frequency on IoT devices. Although such assumptions given the lack of data can only be described as super forecasting, the estimated average attack detection and mitigation in terms of the 5 min from connection to attack timeframe, can be described with different ML algorithms. Comparing the average attack detection and mitigation time in Fig. 1. with multiple algorithms (Nguyen et al. 2019), including Distributed Edge-based Defence (DED), Centralised Fog-based Defence (CFD), Centralised Fog and Cloud-based Defence (CFCD) and SeArch architecture.

---
[1] 365×24×60×.5= 262,800÷5=52,560÷262,800=0.2÷10,000
[2] SonicWall report (SonicWall 2019) captured real-world data from more than one million sensors in over 215 countries with over 140,000 malware samples collected daily.

The average attack detection and mitigation time Fig. 1, shows that although some of the NIDS described earlier, can detect IoT attacks within the 5 min average attack time from the moment of connection, none of the NIDS shown in Fig. 1 can mitigate IoT attacks instantly. Therefore, understanding the risks before they occur is of a significant relevance to preventing severe impact from IoT attacks.

## Conclusion

This study reviewed how different AI methods can be applied for cyber risk analytics in extreme environments, such as exploration of outer space, where AI would need to perform all the tasks, including its own risk assessment. The review confirms that for the integration of AI/ML in risk analytics, we need to adapt the data strategies to collect the appropriate cyber-risk data. With the integration of IoT systems, new types of data streams are becoming available. Such data streams can be collected and analysed with AI/ML algorithms. The survey review in this paper, identified some of the potential impact assessment approaches that can be redesigned for predictive, dynamic and self-adopting cyber risk analytics. The conclusion builds upon the existing approach for categorising (pooling) risk, but presents a quantitative version of the NIST 'traffic lights' system (demonstrated in Table 2), enhanced with multiple risk calculation metrics that calculate the shortfall probability, expected shortfall, VaR and CTE for different cyber risk levels and tail risk under different assumptions (see Table 2).

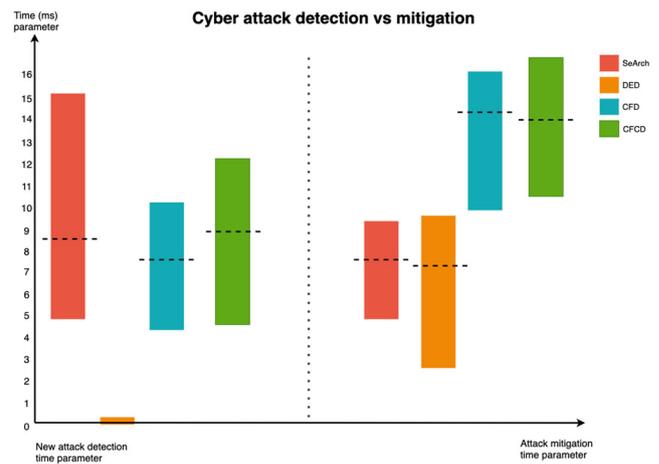

Fig. 1 The average attack detection and mitigation on IoT connected devices

The mathematical formulae present a better understanding of the cost and risk evaluation with multiple risk calculation metrics for different cyber risk levels and tail risk under different assumptions. The value of safety and cyber risk in extreme environments – such as outer space, can be explained in economic terms, where the level of cybersecurity is based on the risk acceptance level and the co-ordination of sufficient protection of the communications networks.

This study presents a mathematical formula for the future cyber risk developments that are reshaping the data analytics of supply and control systems. The mathematical formulae represent an advancement and integration of the NIST 'traffic lights' system and the FAIR-U Tool, though 'pooling' of risk data into calculation metrics, while anonymising data from individual IoT devices.

Secondly, the co-ordination of supply and control systems cyber protection though AI/ML must be reliable to prevent abuse from the AI itself. The mathematical formulae in this article relies on multiple risk calculation metrics, while existing cyber risk assessment approached are designed with individual risk calculation metrics. The integration of multiple risk metrics presents a more robust protection from abuse of individual data intelligence streams.

Thirdly, the predictive cyber risk analytics as presented in the article, are based on different levels of risk intelligence that are 'pooled' into numbers and not presented as individual risk events. Hence, it allows for anonymising the risk data, and after applying the mathematical model, the data is presented into anonymous risk categories.

## Limitations and further research

AI/ML in cyber risk data analytics integrated in the supply chains and control systems would present innovative and cost-effective ways to protect such data. In addition, the AI/ML analysis of the threat event frequency, with a dynamic and self-adopting AI enhanced methodology, would empower the design of a cognition engine mechanisms for predicting the loss magnitude through the control, analysis, distribution and management of probabilistic data. The development of such cognitive engine and its application, would undoubtedly bring multiple benefits and would enable deeper understanding of the impact of cyber risk at the edge. Nonetheless, IoT networks represent complex coupled systems (D. De Roure et al. 2019), that can be described as cyber-physical social machines (Madaan et al. 2018) and social machines (David De Roure et al. 2015) should be observed in practice (Shadbolt et al. 2019). Given that IoT is considered as critical enabler (Lee et al. 2019a) of value creation (Lee et al. 2019b), the findings of this study would probably be best verified when observed in practice.

**Acknowledgments** Eternal gratitude to the Fulbright Scholar Project.

**Availability of data and material** Not applicable

**Authors' contributions** Dr. Petar Radanliev: main author; Prof. Dave De Roure: supervision; Dr. Kevin Page: supervision; Dr. Max Van Kleek, Omar Santos, La'Treall Maddox, Prof. Pete Burnap, and Prof. Carsten Maple: supervision and corrections.

**Funding** This work was funded by the UK EPSRC [grant number: EP/S035362/1] and by the Cisco Research Centre [grant number 1525381].

## Compliance with ethical standards

**Conflicts of interest/competing interests** The authors declare that they have no conflicting and/or competing interests.

**Code availability** Not applicable